\documentclass{svjour2}

\usepackage{graphicx}

\usepackage{rotating}
\usepackage{amssymb}
\usepackage{mathptmx}
\usepackage{textcomp}
\usepackage[numbers]{natbib}

\newcommand{\micro}{\hbox{\textmu}}

\begin{document}

\title{Measurements on Melting Pressure, Metastable Solid Phases, and Molar Volume of Univariant Saturated Helium Mixture}

\author{J. Rysti \and M.~S. Manninen \and J. Tuoriniemi}

\institute{O.V. Lounasmaa (Low Temperature) Laboratory, Aalto University, P.O. Box 15100, 00076 Aalto, Finland\\
\email{juho.rysti@aalto.fi}
}

\date{}

\maketitle

\begin{abstract}
A concentration-saturated helium mixture at the melting pressure consists of two liquid phases and one or two solid phases. The equilibrium system is univariant, whose properties depend uniquely on temperature. Four coexisting phases can exist on singular points, which are called quadruple points. As a univariant system, the melting pressure could be used as a thermometric standard. It would provide some advantages compared to the current reference, namely pure $^3$He, especially at the lowest temperatures below 1~mK. We have extended the melting pressure measurements of the concentration-saturated helium mixture from 10~mK to 460~mK. The density of the dilute liquid phase was also recorded. The effect of the equilibrium crystal structure changing from hcp to bcc was clearly seen at $T=294$~mK at the melting pressure $P=2.638$~MPa. We observed the existence of metastable solid phases around this point. No evidence was found for the presence of another, disputed, quadruple point at around 400~mK. The experimental results agree well with our previous calculations at low temperatures, but deviate above 200~mK.

\end{abstract}

\section{Introduction} \label{Sec:Intro}

At any pressure, liquid $^3$He--$^4$He mixture separates into two phases below $T\sim 0.9$~K, provided enough of both components are present. The so-called dilute phase is rich in $^4$He and has a finite solubility of $^3$He even down to the zero temperature. The $^3$He-rich phase becomes pure $^3$He in the limit $T\to 0$. Increasing the pressure of a phase-separated liquid mixture results eventually in the appearance of a solid phase whereupon the system reaches the melting curve of the concentration-saturated helium mixture. It consists of three or four simultaneous phases. The latter situation defines a quadruple point, which fixes all the intensive thermodynamic variables. The saturated system at the melting pressure is a univariant and its properties are uniquely determined by temperature. It therefore offers the possibility to use it as a thermometric standard in a similar manner as pure $^3$He. The provisional low temperature scale from 0.9~mK to 1~K (PLTS-2000) is based on the melting pressure of pure $^3$He \cite{PLTS2000}. Below its nuclear magnetic ordering temperature $T_\mathrm{N}\approx0.9$~mK, the melting pressure of pure $^3$He is proportional to $T^4$ due to the entropy of solid $^3$He. However, the melting pressure of saturated helium mixture is proportional to $T^2$ because of the entropy of the fermionic normal $^3$He component in the dilute liquid phase. The slope of the melting pressure curve $dP_m/dT$ of the helium mixture system thus remains larger than that of the pure system at very low temperatures and has the potential to offer better resolution there. A differential pressure gauge with $^4$He as reference can be used to increase accuracy \cite{LT24paper,LT25paper}. The slopes of the two systems cross at around 50~\micro K. However, if one uses $^4$He as the reference for the mixture and vacuum for the pure $^3$He, the pressure measurement of the mixture is two orders of magnitude more accurate. Thus the temperature resolutions of the two thermometers cross at around 0.5~mK. Admittedly, operating with a mixture system is more involved than the pure system. However, in some applications, such as the adiabatic melting refrigeration of $^3$He--$^4$He mixtures, this type of thermometer is very natural \cite{TuoriniemiAdiabatic}.

Since the melting pressure of pure $^4$He is lower than that of pure $^3$He, the solid mixture always contains $^4$He. In the limit of zero temperature, the solid phase becomes pure $^4$He \cite{EdwardsBalibar}. In this case the melting pressure is that of pure $^4$He plus an additional contribution due to the osmotic pressure in the dilute liquid \cite{Osmotic}. As temperature is increased, the melting pressure first increases quadratically, since it is mostly determined by the entropy of $^3$He in the degenerate Fermi liquid phases with practically constant solubilities \cite{Solubility}. At higher temperatures, $^3$He dissolves into the solid at larger quantities. This turns the melting pressure down, so that a maximum is obtained at around 300~mK. The melting pressure falls below that of pure $^4$He at about 400~mK and reaches a minimum in the range of 1~K. During increasing concentration of the solid phase, the preferable crystal structure changes from hexagonal close-packed (hcp) of pure $^4$He to body-centered cubic (bcc), which is the same structure as in pure $^3$He. This change occurs at the maximum melting pressure, which then is a quadruple point.

In this paper we report the results of measurements on the melting pressure of the concentration-saturated helium mixture between 10~mK and 460~mK and the observed nucleation pressures for creating new crystals at various temperatures. This work extends the previous experiments, which were measured from below 1~mK to 10~mK \cite{EliasMeltingPressure}. The molar volume of the dilute liquid phase at the melting pressure is given as well. We also report the existence of metastable solid phases around the maximum pressure.

\section{Experimental arrangement}

In the following we describe the experimental setup and the instrument calibrations. Some of the experimental difficulties discussed have nearly negligible effect on the results, but are included for completeness.

The experimental setup on the nuclear refrigerator \cite{Weijun2000} was designed for studies of helium crystals at submillikelvin temperatures \cite{MasaPaper}. However, the experiment described here concerned temperatures only above 10\,mK and thus the nuclear stage with the cell was always thermally connected to the mixing chamber of the dilution refrigerator via an aluminium heat switch.

\begin{figure}
\centering
\includegraphics{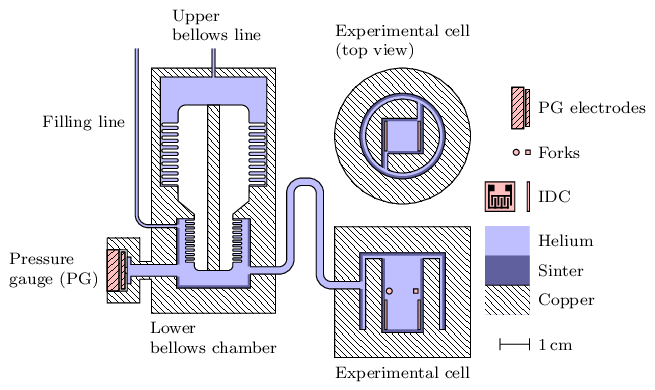}
\caption{(Color online) The experimental setup. The central cuboid volume of the experimental cell was surrounded by an annular volume and contained the interdigital capacitors (IDC) and quartz tuning forks. Pressure in the experimental cell was varied with a hydraulic press (bellows chambers) and measured with a capacitive pressure gauge (PG). The cell was bolted on a copper nuclear demagnetization stage and its walls were covered with sintered silver powder. The solid phase usually nucleated into the central cuboid volume.}
\label{Fig:Cell}
\end{figure}

The sample cell consisted of two volumes, the lower bellows chamber of a hydraulic press and the actual experimental cell, see Fig.~\ref{Fig:Cell}.
Since the melting pressure of $^3$He--$^4$He mixtures has a deep minimum at around 1\,K it is not possible to pressurize helium mixtures at low temperatures above that pressure through a filling line, which will be blocked by solid helium. However, the volume, and thus the pressure of the lower bellows chamber could be altered by regulating the $^4$He pressure in the upper operating bellows chamber. The pressure in the upper bellows was always far below the melting pressure of $^4$He because of the upper to lower bellows compressing area ratio of about $6.7~\mathrm{cm}^2/2.0~\mathrm{cm}^2$.

The experimental cell consisted of a central cuboid volume  ($1\,\textrm{cm} \times 1\,\textrm{cm}\times 2\,\textrm{cm}$) and an outer annular volume which provided a thermal guard. Interdigital capacitors for density measurements and quartz tuning forks for thermometry were located in the cuboid volume. The helium sample was thermalized with porous sintered silver powder on the cell walls. Photographs of the experimental cell are shown in Ref.\,\cite{MasaPaper}.

\begin{figure}
\centering
\includegraphics[width=0.98\textwidth]{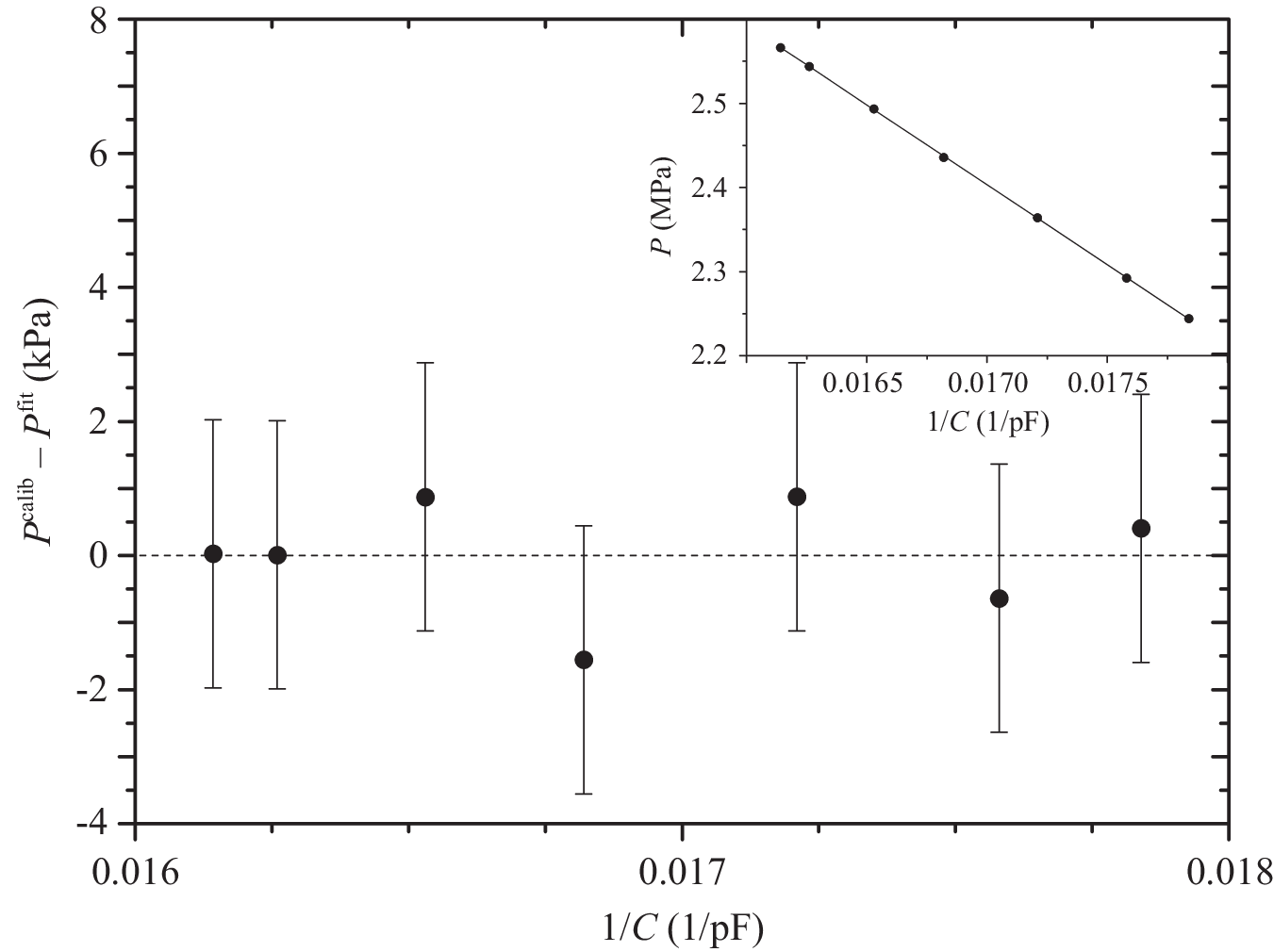}
\caption{Pressure gauge calibration. In the main frame the difference between the calibration data and the resulting fit is given. The error bars reflect the inaccuracy in the reading of the mechanical pressure gauge, which was used for the calibration. The inset shows the absolute data. The capacitive pressure gauge was calibrated against a mechanical pressure gauge at room temperature. The calibration here was performed only to pressures between $P=2.25$~MPa and $P=2.55$~MPa, since the present studies were restricted to the melting pressure. Modest extrapolation to higher pressures, up to $P=2.638$~MPa, is needed to cover the complete range of interest.}
\label{Fig:PGcalibration}
\end{figure}

The capacitive pressure gauge used in these experiments was of a Straty-Adams type made of beryllium copper. It had the sensitivity $dC/dP = 20~\mathrm{pF/MPa}$ in the range of helium mixture melting pressures. It was calibrated seven months earlier using pure $^3$He in the cell with $T=10$~mK and a mechanical pressure gauge at room temperature. The single point calibration accuracy was about 2~kPa. The pressure gauge suffered from slight drifting over time. The scale was therefore further fixed to the zero temperature value of the saturated mixture melting pressure (2.566~MPa \cite{EliasMeltingPressure,Grilly1973}), when used for the present experiments. The capacitances of the $^3$He calibration had to be scaled by 1.0052 to correct for the difference accumulated over the seven months, during which the cryostat was kept mainly below 0.1~K, but was also warmed up to 4~K. This resulted in approximately 11~kPa increase of the reading at the melting pressure, when comparing new and old calibrations. Further minor changes in the pressure gauge were observed during the present measurements after depressurization-pressurization cycles, which were performed several times to increase the amount of $^3$He in the cell. The reading shifted  by an amount between 1~kPa to 4~kPa after each cycle. These were corrected in the data by simply shifting the points so that the values for the solubility saturated system before and after the re-pressurizations matched. The additional small drift with time was of the order of few hundred pascals over the experiment. The pressure gauge calibration is shown in Fig.~\ref{Fig:PGcalibration}. The fitted function is $P/\mathrm{MPa} = 5.6334-190.01(C/\mathrm{pF})^{-1}$.

\begin{figure}
\centering
\includegraphics[width=0.98\textwidth]{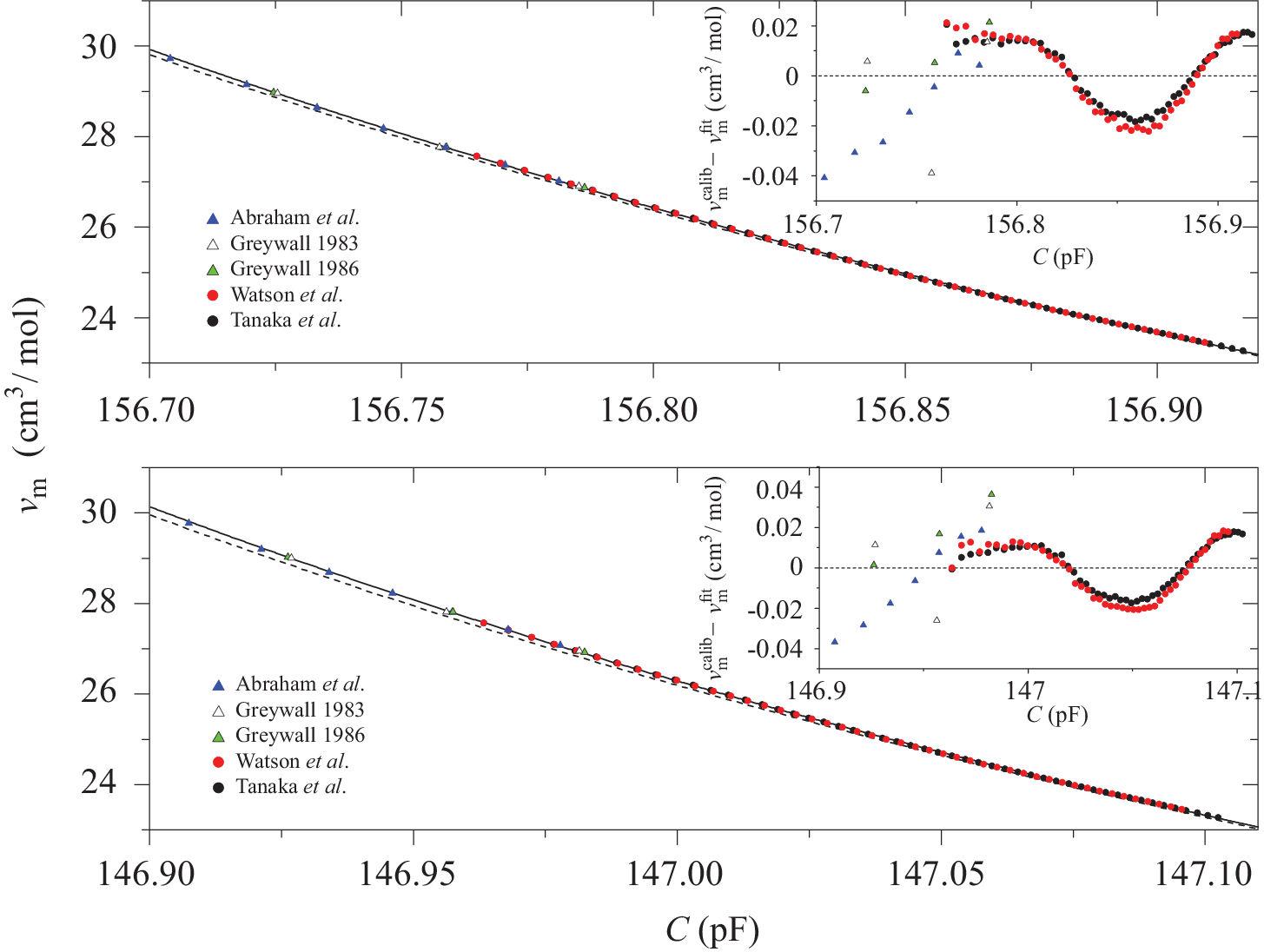}
\caption{(Color online) Interdigital capacitor calibrations. The readings of the two capacitors were fitted to a known functional form obtained from the Clausius-Mossotti relation and the molar polarizability given by Ref.~\cite{MolarPolarizability}. The fit range was between $v_\mathrm{m}=23.3$~cm$^3/\mathrm{mol}$ and $v_\mathrm{m}=29.8$~cm$^3/$mol. The solid curves are the final calibrations using pure $^4$He and a few adjusted $^3$He points. The $^3$He data are indicated by triangles and $^4$He by circles. The dashed curves are the original calibrations done in pure $^3$He. The experimental data at various pressures used for the calibrations were from Refs.~\cite{Watson1969,Tanaka2000,Greywall1983,Greywall1986,Abraham1971}. The insets display the difference between the calibration data and the obtained fits.}
\label{Fig:IDCcalibration}
\end{figure}

Two interdigital capacitors (IDC) on opposite walls of the cell cavity were used to monitor the local helium density. In practice they remained in the dilute liquid phase, except for times when the solid was grown to large enough size. The capacitors were calibrated using pure liquid $^4$He and experimental data from literature \cite{Watson1969,Tanaka2000}. Such calibrations had been performed with pure $^3$He also, when experiments on that were done, but it was observed that between the $^3$He and mixture experiments both capacitors had somewhat inconsistent values at a given nominal helium density. The discrepancy was slightly different for the two IDC's. This was observed when the capacitances of the two capacitors were plotted against each other and results from different liquids were compared. The origin of this effect remains unknown. The difference between the molar polarizabilities \cite{MolarPolarizability} of the two helium isotopes is much too small to account for this and the magnitude of the effect was not exactly the same for the two capacitors. Eventually pure $^4$He over the entire pressure range and a few $^3$He points adjusted to fit the $^4$He data were used for the calibrations. The maximum differences between pure $^3$He and pure $^4$He calibrations were 0.18~cm$^3/\mathrm{mol}$ and 0.12~cm$^3/\mathrm{mol}$ for the two IDC's in the mixture molar volume range. The maximum relative discrepancy was thus 0.7\%. The variation in molar volumes for the two calibrations as functions of capacitances were very similar for both capacitors, see Fig.~\ref{Fig:IDCcalibration}.
This indicates that the majority of the effect was either due to the different helium isotopes used for the calibrations or an inconsistency between the literature values of the two pure liquid molar volumes. The $^3$He calibration gave consistently smaller molar volumes at a given capacitance compared to the $^4$He, with the difference being larger at smaller capacitances (larger molar volumes). The IDC calibrations were performed at 10~mK temperature.

The capacitances $C$ were transformed to molar volumes $v_\mathrm{m}$ by using the Clausius-Mossotti relation
\begin{equation}
\frac{\epsilon -1}{\epsilon +2}=\frac{4\pi \alpha_\mathrm{m}}{3 v_\mathrm{m}},
\end{equation}
experimental data for the molar polarizability $\alpha_\mathrm{m}$ of helium \cite{MolarPolarizability}, and using the expected capacitance for an interdigital capacitor
\begin{equation} \label{Eq:IDC}
C\approx N\epsilon_0 l \frac{\epsilon + \epsilon_\mathrm{s}}{2} .
\end{equation}
In Eq.~(\ref{Eq:IDC}) $N$ is the number of digits, $\epsilon_0$ is the vacuum permittivity, $l$ is the length of the digits, and $\epsilon$ and $\epsilon_\mathrm{s}$ are the relative dielectric constants of the medium and the sapphire substrate, respectively. Two fit parameters for each capacitor, $N\epsilon_0 l$ and $\epsilon_\mathrm{s}$, were used. The fitted values of these parameters were credible based on the actual geometry and materials of the capacitors; $N\epsilon_0 l = 25.674$~pF and $\epsilon_s = 10.391$ for one and $N\epsilon_0 l = 28.198$~pF and $\epsilon_s = 10.062$ for the other. The expected values are $N\epsilon_0 l = 30.1$~pF and $\epsilon_s \approx 10$. The calibrations for the capacitors are represented in Fig.~\ref{Fig:IDCcalibration}. The measurement accuracy was better than 0.1~fF, while the density sensitivity was $dC/dv_\mathrm{m}=-40~\mathrm{fF/(cm^3/mol)}$ in the range of saturated helium mixture molar volumes.

A small but systematic difference between the two IDC's as a function of pressure was observed. This difference corresponded to less than 0.02~cm$^3/$mol over the pressure range from 14~kPa to 3.36~MPa. The IDC's were also observed to feature unexpected temperature dependencies, which differed between the two devices. The temperature dependencies were measured with an empty cell up to a temperature $T=180$~mK and with pure $^4$He at the saturated vapor pressure up to $T=750$~mK, where the molar volume is still practically unchanged. The other capacitor showed some variation even below 10~mK. Above 10~mK its capacitance increased as a function of temperature, reaching a maximum at about 30~mK, where it was 0.4~fF above the 10~mK value. After this the capacitance decreased and leveled off around 300~mK to a value, which was 0.3~fF below the 10~mK reading. For this one the temperature dependence with an empty cell and pure $^4$He did not completely agree, although the general features were similar. The capacitance of the other device decreased monotonically above 10~mK also reaching a stable value at around 300~mK, being 0.5~fF below the 10~mK value. If untreated, these small effects would have changed the molar volumes by 0.02~cm$^3/$mol, or 0.06\%, at most, but they were subtracted from the data assuming the same temperature dependence also in mixtures.

Temperature was measured with a thermal noise thermometer \cite{NoiseThermometer}, which was compared against a commercial calibrated germanium resistance thermometer. The Ge-calibration was checked against the $^3$He vapor pressure (ITS-90) \cite{0026-1394-27-1-002}. The noise thermometer was attached to the dilution refrigerator's mixing chamber.

\section{Results} \label{Sec:Results}

\subsection{Melting pressure}

The melting pressure was measured by performing continuous temperature sweeps with solid and liquid helium in the cell, as well as melting previous solids and nucleating new ones at fixed temperatures. The solid growth and melting at fixed temperatures were controlled by the bellows. A typical re-nucleation was performed by melting the old solid away and decreasing the pressure few tens of kilopascals further. The bellows flow was then reversed and the cell was pressurized at a rate between about $1-2$~kPa/min. Usually the flow to the bellows was halted within a few minutes after nucleation. The appearance of the solid phase was observed as a stop in the increase of pressure as the bellows was pushed at a constant rate. Sometimes the solids were grown to larger sizes, even to cover the IDC's entirely.

\begin{figure}
\centering
\includegraphics[width=0.98\textwidth]{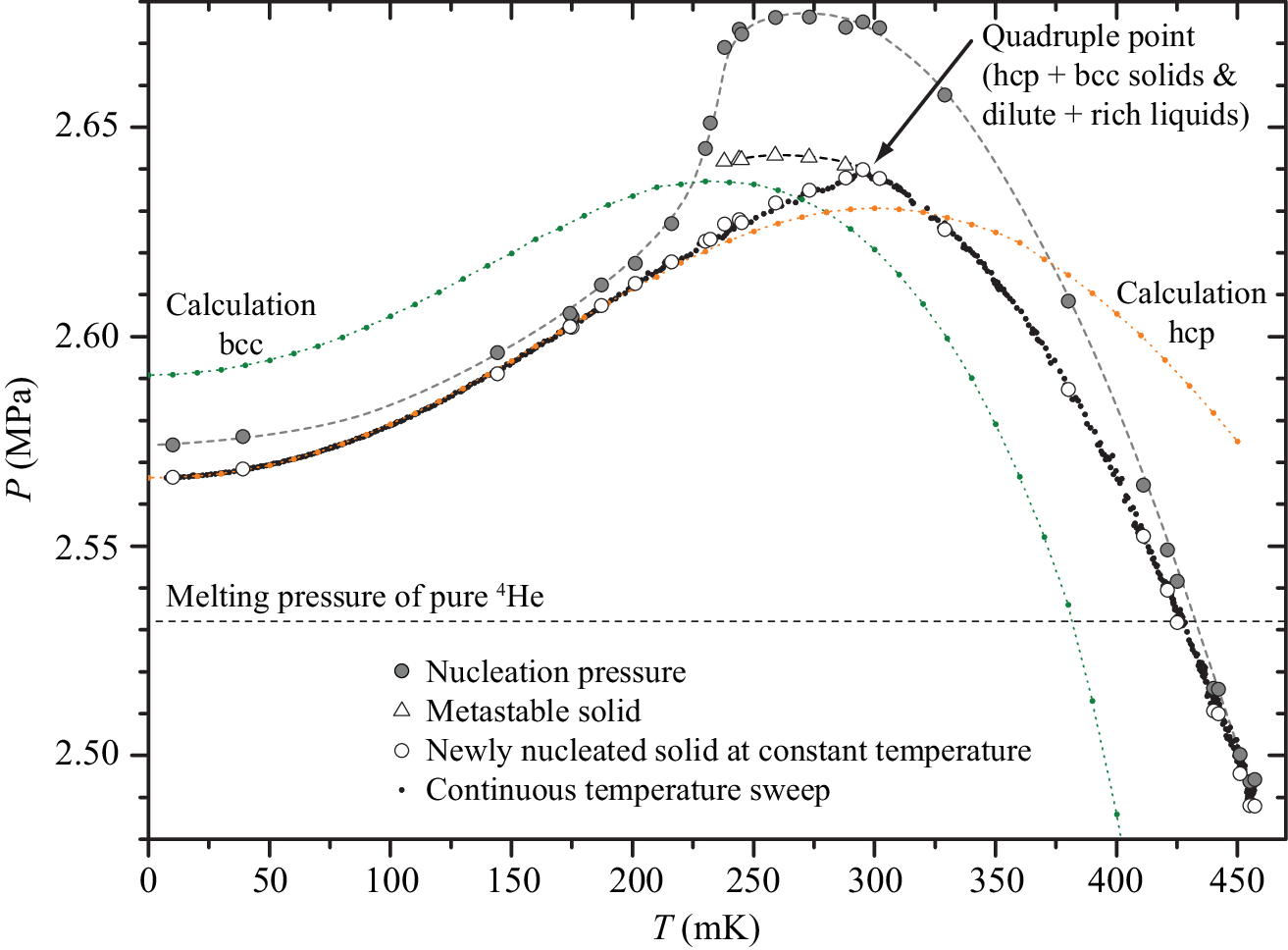}
\caption{(Color online) Melting pressure of the saturated helium mixture as a function of temperature. The open circles represent new crystals nucleated at constant temperatures. The grey circles indicate the required pressures for nucleations. The triangles are for the pressures of metastable solids, which after some time relaxed to the equilibrium pressures. The dashed curves for the nucleation pressure and metastable pressure are guides to the eye. The black dots (forming a nearly continuous line) represent the continuous temperature sweeps. The horizontal dashed line is the melting pressure of pure $^4$He. The dash-dotted lines are the results of earlier calculations for the hcp and bcc crystal structures \cite{MeltPresCalc}. The quadruple point ($T=294$~mK, $P=2.638$~MPa), where hcp solid, bcc solid, dilute liquid, and rich liquid phases coexist, is indicated by an arrow. Below the quadruple point temperature the solid has a hcp structure and above it a bcc structure.}
\label{Fig:Pm}
\end{figure}

The measured melting pressure of the concentration-saturated helium mixture is given in Fig.~\ref{Fig:Pm}.
It shows the results of new nucleations at constant temperatures and the continuous temperature sweeps. The continuous data have been averaged to reduce scatter. The required pressures for nucleations are also indicated. The melting pressure of pure $^4$He is plotted for reference. The two dash-dotted curves are the results of earlier calculations for the hcp and bcc phases \cite{MeltPresCalc}. The melting pressures of new nucleations and the required nucleation pressures are also given in Table~\ref{Tab:Pm}.
\begin{table}
\caption{Melting pressure data of new nucleations. The equilibrium melting pressures $P_\mathrm{m}$ and the required nucleation pressures $P_\mathrm{n}$ are included.}
\label{Tab:Pm}
\centering
\begin{tabular}{lll|lll}
\hline\noalign{\smallskip}
$T$~(mK) & $P_\mathrm{m}$~(MPa) & $P_\mathrm{n}$~(MPa) & $T$~(mK) & $P_\mathrm{m}$~(MPa) & $P_\mathrm{n}$~(MPa) \\
\noalign{\smallskip}\hline\noalign{\smallskip}
10 & 2.567 & 2.574 & 273 & 2.635 & 2.676 \\
39 & 2.569 & 2.576 & 288 & 2.638 & 2.674 \\
144 & 2.591 & 2.596 & 295 & 2.640 & 2.675 \\
174 & 2.602 & 2.606 & 302 & 2.638 & 2.674 \\
175 & 2.602 & 2.605 & 329 & 2.626 & 2.658 \\
187 & 2.608 & 2.612 & 380 & 2.587 & 2.608 \\
201 & 2.613 & 2.618 & 411 & 2.552 & 2.565 \\
216 & 2.618 & 2.627 & 421 & 2.540 & 2.549 \\
230 & 2.623 & 2.645 & 425 & 2.532 & 2.542 \\
232 & 2.623 & 2.651 & 440 & 2.511 & 2.516 \\
238 & 2.627 & 2.669 & 442 & 2.510 & 2.516 \\
244 & 2.628 & 2.673 & 451 & 2.496 & 2.500 \\
245 & 2.627 & 2.672 & 455 & 2.488 & 2.494 \\
259 & 2.632 & 2.676 & 457 & 2.488 & 2.494 \\
\noalign{\smallskip}\hline
\end{tabular}
\end{table}
As seen in Fig.~\ref{Fig:Pm}, the continuous temperature sweeps give results very similar to new nucleations, even though the equilibrium $^3$He concentration in the solid phase varies greatly with temperature. We thus note that the liquid-solid interface remains in equilibrium at all times either because concentration gradients remain trapped inside the solid phase and cause no problems, or for the more unlikely reason of very fast $^3$He concentration relaxation within the solid. This is important for the practical use of the mixture melting pressure as a thermometer. We collected data at about ten-second intervals and thus could not notice effects taking place more rapidly.

\begin{figure}
\centering
\includegraphics[width=0.98\textwidth]{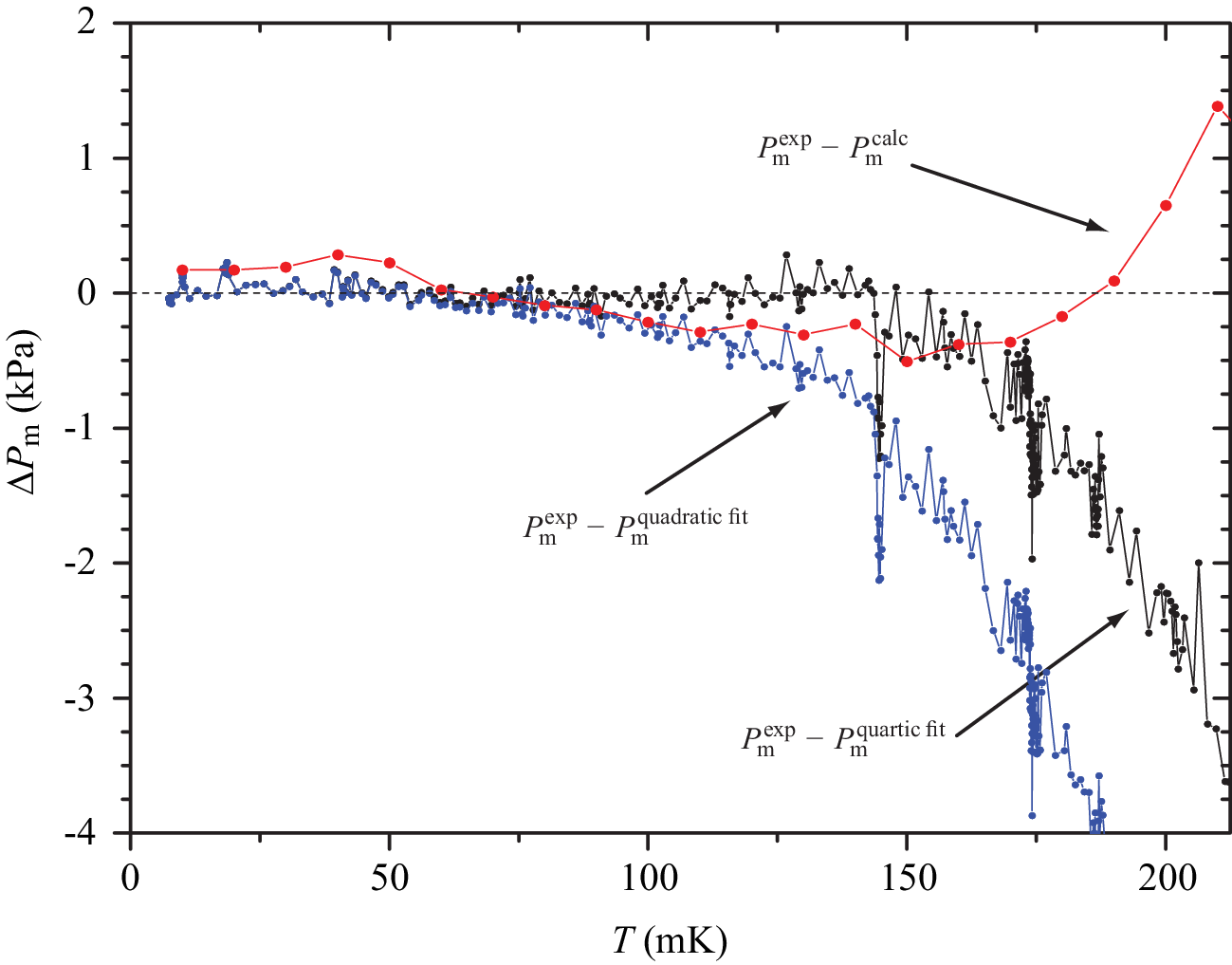}
\caption{(Color online) Difference between the measured melting pressure of the saturated helium mixture and a quadratic fit, a quartic fit, and the calculation of Ref.~\cite{MeltPresCalc}. The quadratic fit has been done using data up to $T=60$~mK and the quartic up to $T=140$~mK.}
\label{Fig:PdiffLowT}
\end{figure}

Low temperature expansion for the melting pressure of the solubility saturated helium mixture can be obtained by fitting a quartic  polynomial to the shown data. Restricting this to temperatures below $T=140$~mK gives $P_\mathrm{m}/\mathrm{MPa} = 2.5663 + 1.2846 (T/\mathrm{K})^2 - 2.0651 (T/\mathrm{K})^4$. The fit was first performed using a second order polynomial below $T=60$~mK and the second order term was kept constant in the quartic fit. Odd powers have been omitted, since this is a Fermi system. The differences between the measured values and the fits are plotted in Fig.~\ref{Fig:PdiffLowT} together with the difference to the calculated pressure of Ref.~\cite{MeltPresCalc}.
In previous experiments at lower temperatures, the value of the quadratic fit parameter, which in our present analysis is 1.28, has been found as 1.1 \cite{LT24paper} or 0.92 \cite{EliasMeltingPressure}. It should be noted that completely different thermometers and pressure gauges have been used in all three experiments. The temperatures cannot be expected to agree to accuracy better than 5\%. The measurements of Ref.~\cite{LT24paper} were not done with the saturated system, but with a concentration of 7\%. Comparison between this and the present experiment is not trivial, because the changing pressure and temperature also change the liquid concentration. From Ref.~\cite{Osmotic} we find that the quadratic parameter is about 5\% larger for the saturated system than for a 7\% system. In Ref.~\cite{EliasMeltingPressure} a very narrow temperature range was used to find the slope above the superfluid transition temperature of pure $^3$He. This increases inaccuracy in the slope, in addition to the possible differences in the temperature and/or pressure scales.

At temperatures below $T=200$~mK the calculated hcp curve of Ref.~\cite{MeltPresCalc} coincides with the experimental one, see Fig.~\ref{Fig:Pm}. Above this temperature the calculated pressures (hcp below the quadruple point and bcc above it) are continually below the measured ones. This includes the entire bcc branch. The hcp-bcc crossing is also predicted to exist at a lower temperature and pressure. This discrepancy is not very surprising. The model interaction potential used to calculate the chemical potential of $^3$He in the dilute liquid phase had been constructed at low temperatures and small concentrations \cite{Potential}. It was known to exhibit some degree of uncertainty at higher temperatures.

\subsection{Metastable states}

Around the maximum melting pressure, metastable solids were observed \mbox{after} nucleations. These are indicated as triangles in Fig.~\ref{Fig:Pm}. The system could remain in these states for several hours before spontaneously relaxing to the equilibrium melting pressures. The relaxation rate could be accelerated by increasing the experimental volume by operating the bellows and thus melting the metastable fraction of the solid. The same pattern continues on the high-temperature side of the maximum pressure as well, but unfortunately the amount of $^3$He in that set of measurements was not enough for a saturated mixture and thus these data points have been omitted from Fig.~\ref{Fig:Pm}. After addition of $^3$He into the cell, no nucleations were performed around the maximum, as the attention was toward the higher-temperature region. Presumably the metastable states are solids with different structures from the equilibrium states. In this case hcp instead of bcc or vice versa.

\begin{figure}
\centering
\includegraphics[width=0.98\textwidth]{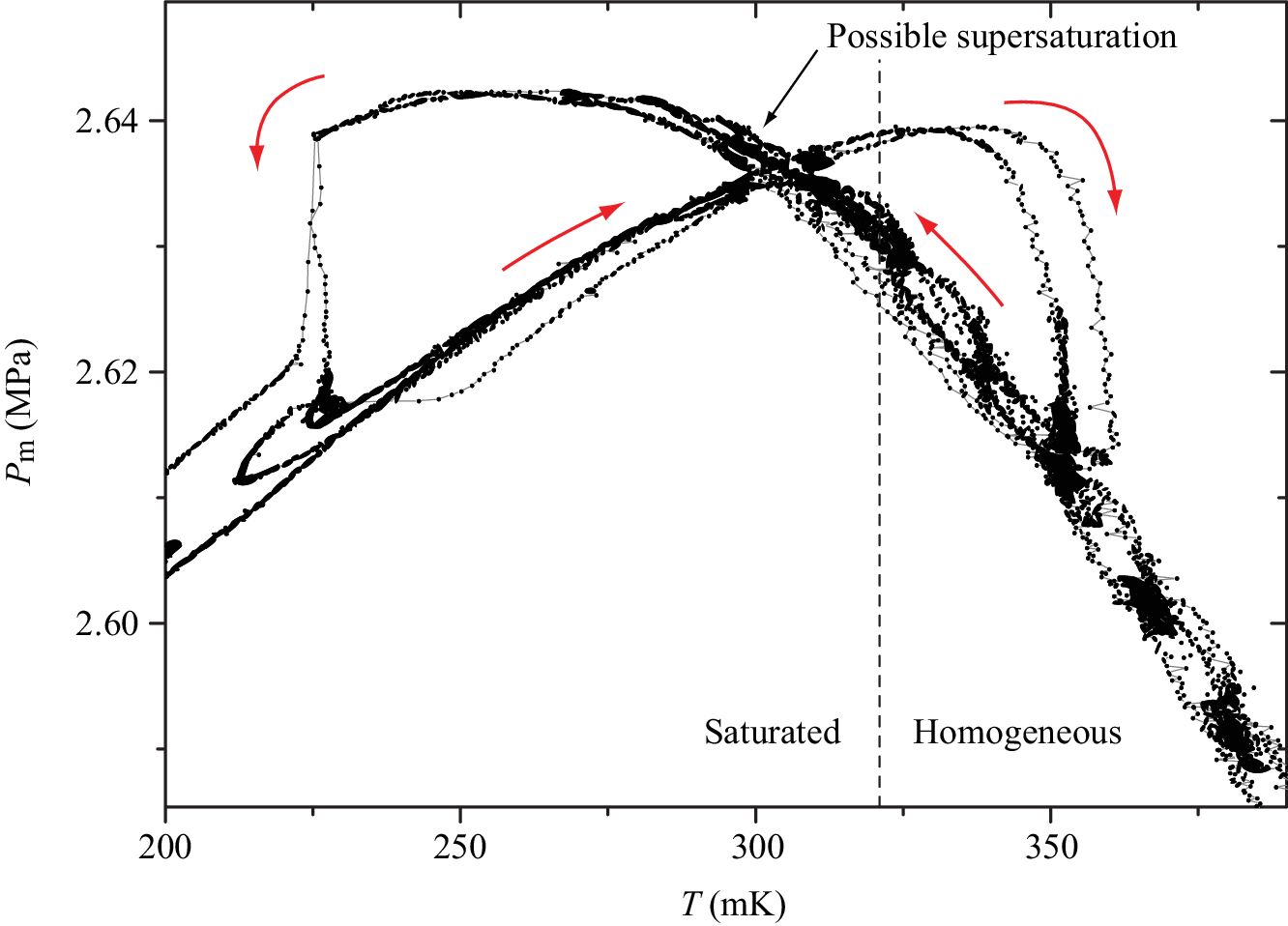}
\caption{(Color online) Melting pressure data obtained by continuously changing the temperature with solid mixture in the cell. No averaging has been performed on these data. The temperature gauge used to plot this figure was a carbon resistor, whose calibration was not as accurate as that of the thermal noise thermometer, but which offered readings more frequently. The temperature scale thus differs slightly from Fig.~\ref{Fig:Pm}. (Mind also that unlike the noise thermometer this resistor was located inside the dilution refrigerator mixing chamber and during temperature changes the temperature inside the mixing chamber is not exactly the same as outside.) Two different phases cross at around $T=300$~mK. The arrows indicate future in time, when reaching the metastable states. Some of the data to the right of the crossing point are for a system with homogeneous liquid, as there was not enough $^3$He in the system to maintain saturation. The dashed line separates the saturated and homogeneous equilibrium systems.}
\label{Fig:PmCrossing}
\end{figure}

The metastable states were also seen in the temperature sweeps. Starting from one side of the maximum, the system would continue on the metastable upper pressure to the other side before collapsing to the lower equilibrium value. This is depicted in Fig.~\ref{Fig:PmCrossing}.
The metastable solids extended further to the lower temperature side than the higher, but due to the steeper descent of the pressure on the high temperature side, the maximum pressure difference was about the same. Since the amount of $^3$He in this case was not enough for the saturated mixture above the hcp-bcc crossing point ($T\gtrsim300$~mK), it is possible that the system entered a supersaturated state upon cooling of the non-saturated mixture. The metastable behavior demonstrated in Fig.~\ref{Fig:PmCrossing} is a clear indication that the crossing pressure is the crossing point of two different phases; in this case the crossing of hcp and bcc solids. This point occurs at the maximum equilibrium pressure, which according to our measurements is $P = 2.638$~MPa at $T=294$~mK. This is the quadruple point consisting of dilute and rich liquid phases and the two solid phases. Calculations by Edwards and Balibar placed this quadruple point at $T=283$~mK and $P=2.63$~MPa \cite{EdwardsBalibar}, measurements of Lopatik at $T=0.28$~K and $P=2.63$~MPa \cite{Lopatik}, van den Brandt \textit{et al.} at $T=0.30$~K and $P=2.63$~MPa \cite{Brandt1982}, and Tedrow and Lee at $T=0.37$~K and $P=2.63$~MPa \cite{Tedrow}. The pressures have been very consistent between different authors, but there has been some discrepancy in the temperature. Our obtained pressure is slightly higher than that of the others, but the temperature fits in between earlier measurements.

\begin{figure}
\centering
\includegraphics[width=1\textwidth]{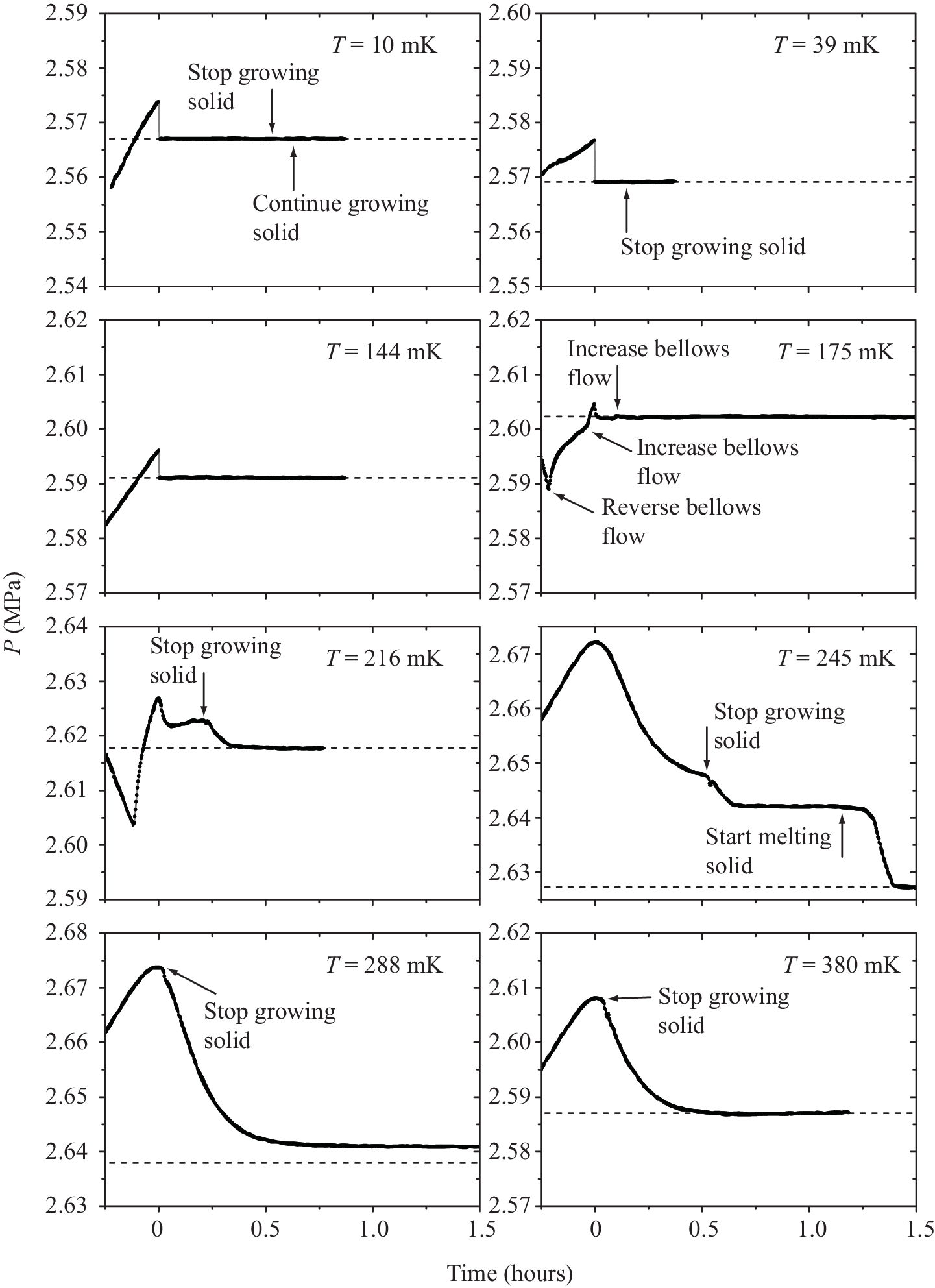}
\caption{Nucleations of solid from helium mixture at eight temperatures. The time scale is the same for all figures and the zero moment represents the initial nucleation. Each figure displays a pressure range of 50~kPa. Metastable solids are seen in the figures for temperatures $T=245$~mK and $T=288$~mK. The horizontal dashed lines in each plot indicate the equilibrium melting pressure. For the $T=288$~mK solid, the system reached the equilibrium state 24 hours after the initial nucleation.}
\label{Fig:Nucleations}
\end{figure}

Some of the nucleation events are plotted in Fig.~\ref{Fig:Nucleations}.
At low temperatures the nucleations were quite fast. The relaxation was markedly slower at higher temperatures. Increasing $^3$He concentration in the solid phase at higher temperatures may have been a contributing factor to this apparent lag in the solid relaxation. More $^3$He must flow from the liquid phases to the solid at higher temperatures. Metastable solids are seen in these figures at temperatures $T=245$~mK and $T=288$~mK. The solid at $T=288$~mK was left on its own and the pressure began to drop from the metastable state about 19 hours after the initial nucleation event. The system reached the equilibrium pressure, which is indicated by a horizontal dashed line, five hours later.

The locations of nucleations in the cell could not be determined most of the time as the solid did not nucleate on the capacitors. This is slightly surprising, since the electric fields of the capacitors should ease nucleation. Above $T=180$~mK the solid nucleated on an IDC only once (at $T=230$~mK). At low temperatures the solid always nucleated on one of the capacitors and if that particular capacitor was not in use, nucleation occurred on the other IDC. However, the nucleation process seemed not to be related directly to the measurement AC voltage nor to the provisionally applied DC voltage, but rather to the on and off switching of the measurement voltage. The capacitors had to be measured alternately, because the two bridges used the same measurement frequency and there was crosstalk between them if active simultaneously. It seems, though, that the nucleations always occurred inside the main sample volume, since reasonable amount of solid growth brought it into the view of the capacitors each time this was done. The location of the pure $^3$He phase could not be observed. If it existed in a volume outside the main experimental cavity, some $^3$He currents in the capillaries must have occurred during growth or melting of the solid.

\begin{figure}
\centering
\includegraphics[width=0.98\textwidth]{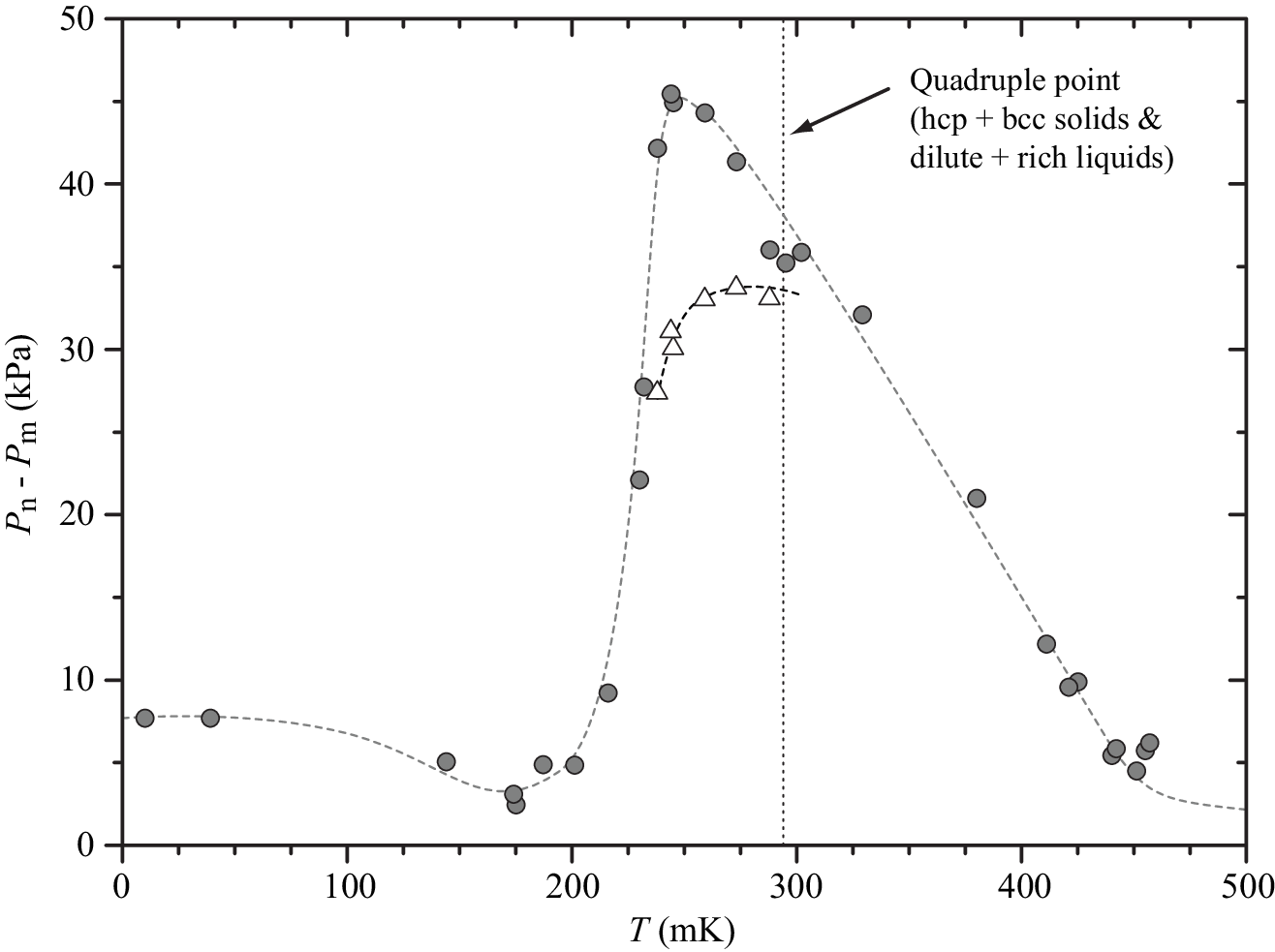}
\caption{Difference between the nucleation pressure and melting pressure. The circles are for the equilibrium pressures and the triangles for the metastable solids. The dashed curves are guides to the eye. The vertical dashed line indicates the quadruple point ($T=294$~mK), where dilute and rich liquids and hcp and bcc solids coexist.}
\label{Fig:OverP}
\end{figure}

Fig.~\ref{Fig:OverP} depicts the required nucleation overpressure relative to the melting pressure.
The dashed line is a guide to the eye. The difference in the pressures peaks at about 250 mK, in the neighborhood of the hcp-bcc crossing (294 mK). At its highest value, the difference is almost 50~kPa. The maximum occurs in the range, where the metastable solids were created upon nucleation. The strong dependence of the nucleation overpressure on temperature is very prominent, but remains without explanation at this time. The effect of different pressurization rates on the required overpressures was not studied systematically. The rates between different nucleations varied more than a factor of two without any noticeable change in the nucleation overpressure.

A solid solution of helium isotopes can experience phase separation into a $^3$He-dilute and $^3$He-rich phases similarly as liquid mixtures. The calculations presented in Ref.~\cite{MeltPresCalc} indicated that the bcc solid at the melting pressure might undergo a separation into dilute and rich phases around $T=380$~mK. This would constitute another quadruple point. Vvedenskii reported having observed this at $T=380$~mK and $P=2.60$~MPa \cite{Vvedenskii}. Tedrow and Lee observed a pressure drop and warming at $T=0.25$~K while cooling, which was attributed to supercooling of the system \cite{Tedrow}. On warming they had a kink in pressure and brief cooling at $T=0.37$~K. Edwards and Balibar concluded from their calculations that the quadruple point does not quite exist \cite{EdwardsBalibar}.

We searched for such possible quadruple point (rich and dilute liquids and bcc solids) in the temperature range of the experiment. The quadruple point would most likely result in a small kink in the melting pressure curve. This, however, may not be very easy to notice. Therefore, we tried to find the quadruple point by detecting the rate of $^3$He accumulation into the solid. Essentially we grew solids at various temperatures so long as to deplete the pure $^3$He liquid phase. This was detected with the capacitors in the dilute liquid phase. Nothing indicated a sudden change in the rate of $^3$He accumulation into a growing solid at any temperature. We must therefore conclude that according to our measurements, the disputed quadruple point does not exist. It is quite possible that the prediction of its existence in our calculations was due to the already-mentioned inaccuracy of the model at those temperatures.

\subsection{Molar volume}

\begin{figure}
\centering
\includegraphics[width=0.98\textwidth]{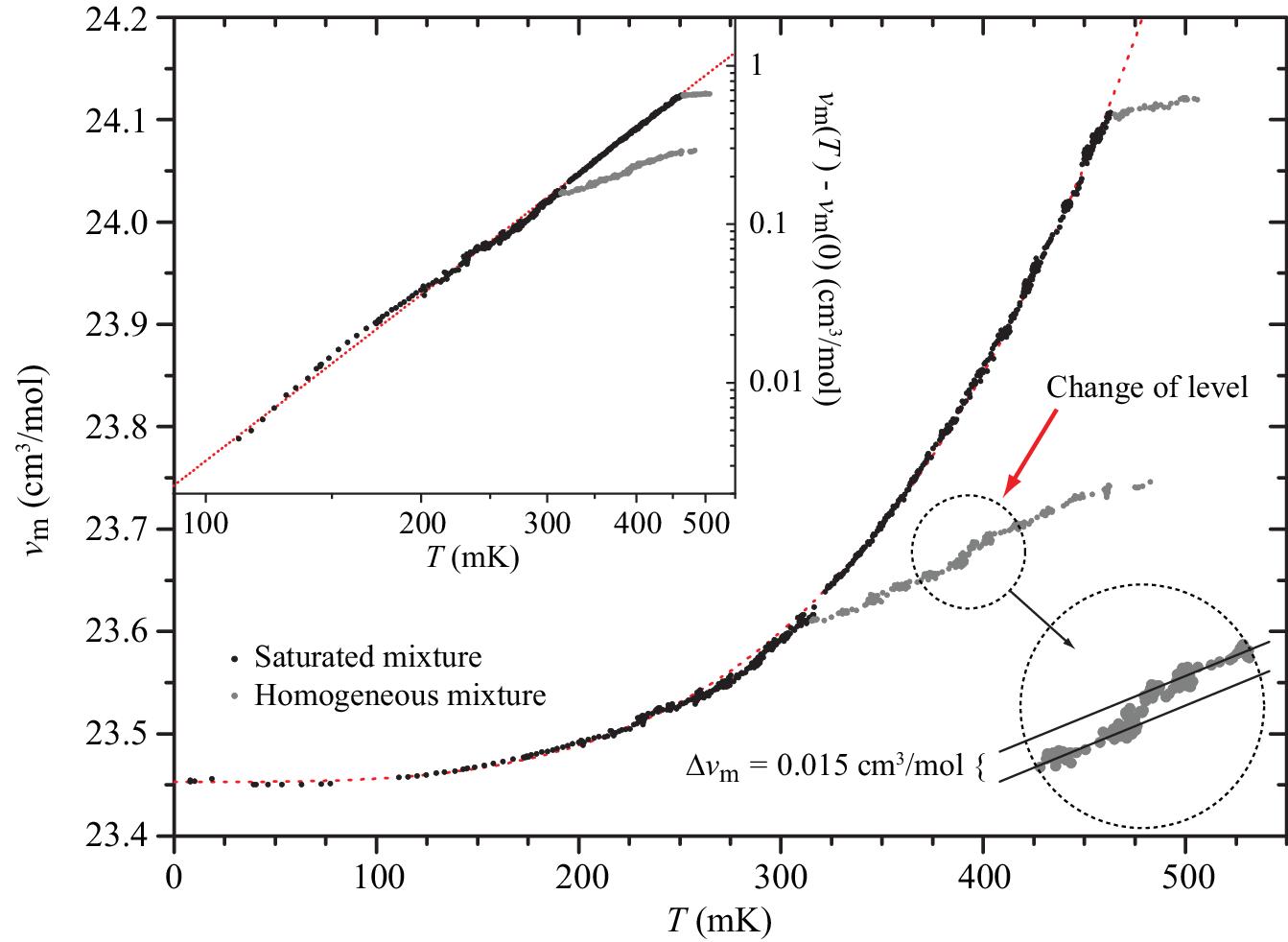}
\caption{(Color online) Molar volume of liquid helium mixture at the melting pressure as a function of temperature. The figure includes the data of both capacitors. The black symbols represent saturated mixture, whereas the grey points are for homogeneous mixtures with different concentrations. Somewhat crude estimate gives $x^\mathrm{L}=15$\% for the lower and $x^\mathrm{L}=24$\% for the higher concentration (see text). The inset shows the same data as a log-log plot with the zero-temperature value subtracted. The dotted curve and line are the power law fit to the data. A reproducible transition to a different level of molar volume was observed in the more dilute homogeneous liquid at around $T=400$~mK. This feature is shown in more detail in the dashed circle, which zooms in the region of interest. The two levels are indicated by the black lines.}
\label{Fig:Vm}
\end{figure}

The measured molar volume of the saturated dilute liquid phase is given in Fig.~\ref{Fig:Vm}.
The black symbols represent the saturated mixture and the grey dots are for homogeneous liquids. Adjacent points have been averaged to reduce scatter. At higher temperatures, data with larger crystals have been omitted, because then systems with large amount of solid have a lower homogenization temperature. This is because the crystal contains relatively more $^3$He than the liquid. We performed measurements with two different $^3$He quantities, which is seen as two different homogenization temperatures ($T=311$~mK and $T=460$~mK). $^3$He was added by pressurizing the cell with about $7\%/10\%/15\%$ mixtures and then decreasing the pressure multiple times at low temperature, which resulted in accumulation of $^3$He into the low temperature cell. Therefore, the overall concentrations were not known to very good precision. Assuming a quadratic temperature dependence for the liquid saturation concentration and using values from Ref.~\cite{Solubility}, we find the total concentrations to have been $x^\mathrm{L}\sim 11$\% and $x^\mathrm{L}\sim 15$\%. The actual concentrations were undoubtedly larger, since the following terms in the polynomial expansion for $x^\mathrm{L}$ are positive. We will return to the issue of the true $^3$He content below.

The molar volume obeys the power law $v_\mathrm{m} \propto T^{3.5}$ remarkably well over the entire temperature range of the saturated system. This is demonstrated in the inset of Fig.~\ref{Fig:Vm}, which shows a log-log plot of the data. The fitted power law, which is $v_\mathrm{m} = 23.453 + 9.663~(T/\mathrm{K})^{3.477}$ $\mathrm{cm^3/mol}$, is also shown in the figure. Some of the data between about 230~mK and 300~mK, where the molar volume slightly deviates from the power law, include readings from the metastable solids. The molar volume of an equilibrium state was about 0.01~$\mathrm{cm}^3/\mathrm{mol}$ larger than the volume of the metastable state at that temperature.

The molar volume of a dilute liquid helium mixture $v_\mathrm{m}^\mathrm{L}$ is often written in the form
\begin{equation} \label{Eq:MolarVolume}
v_\mathrm{m}^\mathrm{L} = v_{40}^\mathrm{L}(1 + \alpha x^\mathrm{L}) ,
\end{equation}
where $v_{40}^\mathrm{L}$ is the molar volume of pure liquid $^4$He, $x^\mathrm{L}$ is the $^3$He concentration, and $\alpha$ is the so-called Bardeen-Baym-Pines parameter, or excess volume parameter. Using our measured molar volume, we can evaluate $\alpha$ in the zero temperature limit. For this we need the molar volume of $^4$He extrapolated to the melting pressure of mixtures ($v_{40}^\mathrm{L}=23.160~\mathrm{cm^3/mol}$ at $P=2.566~\mathrm{MPa}$) \cite{Watson1969,Tanaka2000} and the solubility at the melting pressure $x^\mathrm{L} = 8.12$\% \cite{Solubility}. The result is $\alpha = 0.15\pm0.04$, which corresponds well with the extrapolated value $\alpha \approx 0.16$ of Watson \textit{et al.} \cite{Watson1969}. We could use our data to find $\alpha$ over the entire measured range of melting pressure, but the solubility of $^3$He in $^4$He at the melting pressure has not been determined as a function of temperature beyond some tens of millikelvins. We can, however, turn this consideration around and find better estimates for the maximum liquid concentrations in our experiments by using the molar volume data. We now use the obtained zero-temperature value of $\alpha$ and neglect its temperature dependence as a minor effect \cite{PhysRev.188.309,PhysRevB.67.094503}. We further extrapolate the pressure dependence of this parameter from the data of Watson \textit{et al.} \cite{Watson1969}. We find $x^\mathrm{L}=15$\% for the more dilute system and $x^\mathrm{L}=24$\% for the more concentrated one. These are in accordance with the supposition that the true concentrations were higher than obtained by assuming just a quadratic temperature dependence for the solubility.

From the slopes of the homogeneous cases $dv_\mathrm{m}/dT$, Eq.~(\ref{Eq:MolarVolume}), and the measured melting pressure we can compute $d x^\mathrm{L}/dT$. The result at $T=325$~mK is $dx^\mathrm{L}/dT \approx 3\%/\mathrm{K}$ and at $T=460$~mK it is $dx^\mathrm{L}/dT \approx -32\%/\mathrm{K}$. This means that at the lower temperature a growing solid takes more $^4$He compared to the concentration of the dilute liquid phase and thus tends to slightly concentrate the liquid. At the higher temperature the situation is reversed, and the solid takes most of the $^3$He, diluting the liquid. This is exactly what is expected, since the equilibrium concentration in the solid phase increases much faster as a function of temperature compared to the liquid \cite{MeltPresCalc}.

On the lower homogeneous branch at approximately $T=400$~mK a transfer to a different level of molar volume is seen. It was reproducible when going both up and down in temperature and both capacitors, which were completely independent of each other, gave the same result. The significance of this effect is not clear.

\section{Discussion}

We measured the melting pressure of the saturated helium mixture between $T=10$~mK and $T=460$~mK. For helium mixtures to be used as a thermometric standard, further measurements would be desirable for a consistency check, preferably with a differential pressure gauge using pure $^4$He as reference.

The effect of the equilibrium crystal structure changing between hcp and bcc was clearly observed in the melting pressure data producing a quadruple point on the curve. We found no evidence for the existence of another quadruple point consisting of dilute and rich liquid phases and dilute and rich bcc phases, which had been indicated by an earlier calculation and one other experiment.

Some features were observed, which are merely reported without explanation at this time. The nucleation overpressure has a strong dependence on temperature, attaining its maximum in the region of the maximum melting pressure, where the hcp-bcc coexistence point exists. On both sides of the maximum pressure, long-living metastable solids were first nucleated before an equilibrium pressure was reached. Finally, a small feature in the molar volume of homogeneous dilute liquid phase with coexisting solid phase seems to exist at around $T=400$~mK for $x^\mathrm{L}\approx 15$\%.

\begin{acknowledgements}
This work has been supported in part by the EU 7th Framework Programme (FP7/2007-2013, Grant No. 228464 Microkelvin) and by the Academy of Finland through its LTQ CoE grant (project no. 250280). We also acknowledge the National Doctoral Programme in Materials Physics for financial support. We thank A.~Sebedash and I.~Todoshchenko for useful discussions.
\end{acknowledgements}

\bibliographystyle{apsrev4-1}
\bibliography{refs}   

\end{document}